\documentclass[prb,aps,twocolumn,epsfig,eqsecnum]{revtex4}
\usepackage{epsfig}
\usepackage{bm}
\usepackage{pstricks}

\usepackage{amsfonts}
\usepackage{amsthm}
\usepackage{amsmath}
\usepackage{amssymb}

\newlength{\myL}

\newcommand{\beq}{\begin{equation}}
\newcommand{\eeq}{\end{equation}}
\newcommand{\bea}{\begin{eqnarray}}
\newcommand{\eea}{\end{eqnarray}}

\def\tit#1#2#3#4#5{{#1}{\bf #2}, #3 (#4)}

\def\rmp{Rev.\ Mod.\ Phys.\ }

\def\prl{Phys.\ Rev.\ Lett.\ }

\def\prb{Phys.\ Rev.\ B\ }

\def\jpco{J.\ Phys.\ Cond.\ Mat.\ }

\def\jpsj{J.\ Phys.\ Soc.\ Jpn.\ }

\begin{document}

\title{Slave spin cluster mean field theory away from half-filling: Application to
the Hubbard and the extended Hubbard Model}

\author{S. R. Hassan$^{1}$, L. de' Medici$^{2}$} 

\affiliation{$^1$Department de Physique,Universit\'e de Sherbrooke,
Qu\'ebec, Canada J1K 2R1}
\affiliation{$^2$Department of Physics and Center for Materials Theory, Rutgers University, Piscataway, NJ 08854, USA}

\begin{abstract}

A new slave-spin representation of fermion operators has recently been proposed for the half-filled Hubbard model. We show that with the addition of a gauge variable, the formalism can be extended to finite doping. The resulting spin problem can 
be solved using the cluster mean-field approximation. This approximation takes short-range correlations into account by exact diagonalization on the cluster, whereas long-range correlations beyond the size of clusters are treated at the mean-field level. In the limit where the cluster has only one site and the interaction strength $U$ is infinite, this approach reduces to the Gutzwiller approximation. There are some qualitative differences when the size of the cluster is finite. We first compute the critical $U$ for the Mott transition as a function of a frustrating second-neighbor interaction on lattices relevant for various correlated systems, namely the cobaltites, the layered organic superconductors and the high-temperature superconductors. For the triangular lattice, we also study the extended Hubbard model with nearest-neighbor repulsion. In additionto a uniform metallic state, we find a $\sqrt(3) \times \sqrt(3)$ charge density wave in a broad doping regime, including commensurate ones. We find that in the large $U$ limit, intersite Coulomb repulsion $V$ strongly suppresses the single-particle weight of the metallic state.

\end{abstract}

\maketitle
\section{Introduction}
The theoretical description of strongly correlated systems, such as 
high temperature superconductivity, heavy fermions, and ultra cold atoms in optical lattices, etc.,
poses major challenges in field of the condensed matter physics. These are all systems where the strength
of the electron-electron interaction is comparable to or greater 
than the kinetic energy of the electrons, i.e., any theory based on a 
perturbative expansion around the non interacting limit is at least questionable.
The non perturbative nature of the problems adds extreme difficulty to theoretical
tools describing these systems. In recent years, several radically new and reliable
non perturbative approaches to the problem of strong correlations have been developed
such as Dynamical Mean-Feild Theory(DMFT)\cite{Georges}, Dynamical Cluster approximation\cite{DCA}, 
Cluster-DMFT\cite{CDMFT}, Variational Cluster 
Approximation (VCA)\cite{VCPT}, Two-Particle Self-Consistent Approach (TPSC)\cite{Vilk}; 
these new approaches have led to substancial progress in our understanding of these systems. 

  Some other non-perturbative semi-analytic approaches based on the idea of slave-variable 
representations  of correlated fermions have also been devised and have been used for decades now, 
in order to perform non-trivial approximations on many-body models. In this respect slave-bosons have 
been particularly successful. Their formulation in the limit of infinite correlation between 
the electrons\cite{ColemanSB} can be systematically introduced as a saddle point approximation 
plus corrections,  and has lead to much insight in the physics of the strongly correlated systems, 
most notably of heavy fermions. The alternative formulation that can treat finite interaction 
strength\cite{KotliarRuckenstein} cannot be controlled as a saddle point, but it turns 
out to be a very practical implementation of the Gutzwiller approximation. It has been 
generalized to many-orbital models\cite{Lechermann_NewSB} and succeeded in capturing the 
essential of quasiparticle physics stemming out from the competition between interactions 
and delocalization energy. High energy features can be also studied from fluctuations around 
this mean-field.

The main limitation of this last formulation is the fact that the number of slave-variables 
increases exponentially with the number of degrees of freedom in the mean field, making 
multi-orbital or cluster mean-field quickly intractable.

A different approximation, based on quantum rotors as slave-variables\cite{Florens_rotlong} has 
been devised that is much more economical since it introduces only one slave variable per site, dual to 
the total on-site charge. Still this technique can only be used correctly at half-filling and cannot address 
orbital-dependent observables or magnetic properties of the system. It has been nevertheless successfully applied to 
cluster men-fields recently\cite{Paramekanti}. Also an extension of this technique controlled by large 
degeneracy limits has revealed itself very powerful as an impurity solver\cite{Florens_Rotors}.

    Recently, a new representation of fermion operators that instead uses
quantum spins as slave variables was proposed to study the multi-band Hubbard model 
at half-filling\cite{Luca}. In this paper, we generalize this 
representation away from half-filling and apply it to study  
Mott transition on the different lattices and  the charge denstity wave 
(CDW) transition on the triangular lattice. The Hubbard model plays the 
role of a standard model for correlated fermions on a lattice; it 
contains the band kinetic energy and the local on-site interaction.  
In order to study CDW, the Hubbard model was extended to include an 
intersite electron-electron interaction (V). This leads to the so called  
extended Hubbard model (EHM). Recently, EHM and its variant 
on the triangular lattice have been extensively studied in the context of 
cobaltates\cite{Watanabe, Motrunich1,Motrunich2,Baskaran,Brijesh,Lee1,Zheng,GTPSC}. 


 The Hamiltonian for the extended Hubbard model (EHM) on  
a two dimensional lattice  with sites labeled by $i$ is 
\begin{align}
H&=\sum_{<ij>}-t(d^{\dag}_{i\sigma}d_{i\sigma}+hc)-\mu\sum_{i}n{_i}+\frac{U}{2}\sum_{i}(n_{i}-1)^2 \nonumber \\
&+V\sum_{<ij>}(n_{i}-1)(n_{j}-1),
\end{align}
where $\mu$, $t$, $U$ and $V$ are the chemical potential, the nearest-neighbor hopping amplitude, 
the on-site local  interaction U and nearest-neighbor interaction V, respectively, 
$d_{i\sigma}$ ($d^{\dag}_{i\sigma}$) destroys (creates) an  electron on site $i$ with spin 
$\sigma$, $<i,j>$ denotes that the sum is over nearest neighbors only and the 
number operator is $n_{i}\equiv\sum_{\sigma} d^{\dag}_{i\sigma}d_{i\sigma}.$ 

In order to treat this problem in the simplest approximation that is capable to lead to insight 
on the physics of short-range correlations, we employ a cluster mean-field approximation based on 
the slave-spin representation. We have recently shown that cluster mean-field approximation
for bosons successfully describe the supersolid phase and phase diagram of bosons on 
triangular lattice\cite{Hassan}.

In the following section, we introduce the method.  In particular we introduce the gauge needed to its extension off filled regimes. Sec. III presents the results on 
the Hubbard model, and Sec. IV those on the Extended Hubbard model. We then summarize 
and conclude. Appendices contain various technical details such as the choice of gauge 
and the infinite U limit.

\section{ Slave-spin mean field theory}

Slave-spin mean-field theory\cite{Luca} is the ideal bridge between  
the slave-variable techniques mentioned in the introduction, when taken at the mean-field 
level, in that it provides full insight in multi-obital and cluster 
cases, but still remains the most economical way to do this, since it 
introduces only one slave variable (a spin-$1/2$) for every degree of 
freedom in the mean-field cluster. In practice, for a single-site mean field 
of a one-band model, two slave-spins (one for spin-up electrons and one for 
spin-down electrons) are used, whereas for an N-orbital local mean-field or 
a N-site cluster mean-field of a one-band model the number raises only to $2N$.
Each slave spin increases the size of the Hilbert space by a factor of two. 
The gain is thus enormous compared with slave-boson representations 
because the number of bosons there grows exponetially.

Where detailed comparison have been performed one finds, as discussed below, that the slave-spin mean-field reproduces the 
results of the Gutzwiller approximation, even if a precise mapping has not yet 
been rigorously derived.


\subsection{Slave-spin representation for arbitrary filling}

In the slave-spin representation, we map the original local Hilbert space of the problem onto a larger local Hilbert space
that contains as many fermionic degrees of freedom (named $f_{i\sigma}$)  as the original plus the same number of 
spin-$1/2$ quantum variables, one for each $f_{i\sigma}$\footnote{It is worth noting that the auxiliary spin has nothing to do with thephysical spin, that is treated here just as a label. In fact a slave-spin variable is introduced for every fermion species, taking in account the fermion spin multiplicity, so that slave-spins are also labeled with a spin index i.e. $S^z_{i\sigma}$}. We then associate to every state of the original physical space one of the states in this larger space by using the correspondence:
\begin{equation}
\vert n^d_{i\sigma}=1\rangle \Longleftrightarrow \vert n^f_{i\sigma}=1, \, S^z_{i\sigma}=+1/2\rangle ,
\end{equation}
\begin{equation}
\vert n^d_{i\sigma}=0\rangle \Longleftrightarrow  \vert n^f_{i\sigma}=0, \, S^z_{i\sigma}=-1/2\rangle.
\end{equation}
In words, when a local orbital and spin state is occupied then the corresponding slave-spin is "up" and if it is empty the slave-spin is "down". With these one-particle states one construct the many-particle states as usual.

The enlarged local Hilbert space contains also unphysical states 
such as $\vert n^f_{i\sigma}=0, S^z_{i\sigma}=+1/2\rangle$ and $\vert n^f_{i\sigma}=1, S^z_{\i \sigma}=-1/2\rangle$. 
These unphysical states are excluded if the following local constraint is enforced at each site and for each $\sigma$:
\begin{equation}
f^{\dag}_{i\sigma}f_{i\sigma}= S^z_{i\sigma}+\frac{1}{2}.
\label{constraint}
\end {equation}

We then have to map the operators onto operators that act in the enlarged Hilbert space.
The electron number operator is  easily represented by the auxilary fermions number, i.e.,
$n^d_{i\sigma}=n^f_{i\sigma}$, but also by the z component of the slave-spin $n^d_{i\sigma}=S^z_{i\sigma}+1/2$, thanks to the constraint. This allows us to rewrite the density-density interaction terms in the hamiltonian in terms of the spins only:
\begin{align}
H_{int}&=\frac{U}{2}\sum_{i}(\sum_{\sigma}S^{z}_{i\sigma})^2  \nonumber \\
& + V\sum_{<i,j>}(\sum_{\sigma}S^{z}_{i\sigma})(\sum_{\sigma}S^{z}_{j\sigma})
\end{align}

For the non-diagonal operators we generalize the prescription of Ref. \cite{Luca}, i.e.
\begin{equation}\label{eq:old_prescription}
d_{i\sigma} = f_{i\sigma}2S^x_{i\sigma},\qquad d_{i\sigma}^{\dag} = f_{i\sigma}^{\dag}2S^x_{i\sigma}
\end{equation}
(where $f_{i\sigma}$ is the auxiliary fermion annihilation operator) to the more general one
\begin{equation}
d_{i\sigma} = f_{i\sigma}O_{i\sigma},\qquad d_{i\sigma}^{\dag} = f_{i\sigma}^{\dag}O_{i\sigma}^{\dag}
\label{eq:c}
\end{equation}
in which  $O_{i\sigma}$ is a generic spin-$1/2$ operator, i.e.  a $2\times 2$ complex matrix.

Indeed it is easy to determine that the most general form for $O_{i\sigma}$ is
\begin{equation}\label{eq:O}
O_{i\sigma}=\left(\begin{array}{cc} 0 & c_{i\sigma}\\1 & 0 \end{array}\right),
\end{equation}
where $c_{i \sigma}$ is an arbitrary complex number (When $c_{i\sigma}$ is not of unit modulus, there is no problem with anticommutation relations, if they are taken between physical states.), in order for the operator (\ref{eq:c}) to have, in 
the physical states of enlarged Hilbert space, the same effect
as the fermionic operators in the original Hilbert space, i.e.;
\begin{eqnarray}
d_{i\sigma}|n^d_{i\sigma}=0\rangle=0, &&\quad d_{i\sigma}|n^d_{i\sigma}=1\rangle=|n^d_{i\sigma}=0\rangle\nonumber \\
d^{\dag}_{i\sigma}|n^d_{i\sigma}=1\rangle=0,&&\quad d^{\dag}_{i\sigma}|n^d_{i\sigma}=0\rangle=|n^d_{i\sigma}=1\rangle
\end{eqnarray}

The arbitrariness of the complex number $c_{i\sigma}$ is a gauge of our formulation and stems out 
from the fact that different operators can have the same effect in the physical subspace of 
the enlarged Hilbert space, while acting differently on the unphysical states. This difference 
does not have any effect as long as the constraint is treated exactly.  In practice the local constraints 
are enforced via Lagrange multipliers and approximations have to be performed on these and on the Hamiltonian 
in order to solve the model. In these approximations the particular choice of gauge comes into play. 
$c_{i\sigma}$ can indeed be tuned in order to give rise to the most physical approximation scheme, by  
imposing, for instance, that it correctly reproduces solvable limits of the problem, like the non-interacting 
limit. We will see that the correct choice of $c_{i\sigma}$ depends on the average occupation of the local 
state, and is such that it reduces to 1 at occupation $1/2$, so that $O_{i\sigma}=2S^x_{i\sigma}$ and 
the prescription (\ref{eq:old_prescription}) used at half-filling in Ref.\cite{Luca} is correctly 
recovered.

Finally, in the enlarged Hilbert space the Hamiltonian can be written exactly as:
\begin{align}
H=&-t\sum_{<ij>\sigma}O^{\dag}_{i\sigma}O_{j\sigma}f^{\dag}_{i\sigma}f_{j\sigma}-\mu\sum_{i\sigma}n^f_{i\sigma}\\
&+\!\!\! \frac{U}{2}\sum_{i}(\sum_{\sigma}S^{z}_{i\sigma})^2  + V\!\!\sum_{<i,j>}(\sum_{\sigma}S^{z}_{i\sigma})(\sum_{\sigma}S^{z}_{j\sigma}),\nonumber
\end{align}
subject to the constraint  (\ref{constraint}).

\subsection{Mean-field approximation}
An approximation is now introduced, which consists in three main steps: 1) treating the constraint on average, 
using a static and site-dependent (but spin-independent, since we will not investigate here magnetic phases) 
Lagrange multiplier $\lambda_{i}$ 2) decoupling auxiliary fermions and slave-spin degrees of freedom and 
finally 3) treating the slave-spin Hamiltonian in a cluster mean-field approximation (CMFA), that takes into 
account the nearest neighbor correlations induced by $V$.

After the first two steps, the total Hamiltonian can be written as the sum of the following two effective 
Hamiltonians:
\begin{align}
H_{f}=&-t\sum_{<i,j>,\sigma}Q_{ij}f^{\dag}_{i\sigma}f_{j\sigma}+ H.c.
-\sum_{i}(\mu+\lambda_{i})n^f_{i},   \\
H_{s}=&-\sum_{<ij>,\sigma}J_{ij}O^{\dag}_{i\sigma}O_{j\sigma} + H.c. 
+\sum_{i,\sigma}\lambda_{i}(S^{z}_{i\sigma}+\frac{1}{2})\nonumber \\
&+\frac{U}{2}\sum_{i}(\sum_{\sigma}S^{z}_{i\sigma})^2
+ V\sum_{<i,j>}(\sum_{\sigma}S^{z}_{i\sigma})(\sum_{\sigma}S^{z}_{j\sigma}) .
\end{align}
 The parameters $Q_{ij}$ (effective hopping), $J_{ij}$
(slave-spin exchange constant) and $\lambda_{i}$ in these expression are 
determined from the following  coupled self-consistency equations:
\begin{equation}
Q_{ij}=<O^{\dag}_{i\sigma}O_{j\sigma}>_{s},
\end{equation}
\begin{equation}
J_{ij}=t<f^{\dag}_{i\sigma}f_{j\sigma}>_{f},
\end{equation}
\begin{equation}
<n^f_{i\sigma}>_{f}=<S^{z}_{i\sigma}>_{s}+\frac{1}{2},
\end{equation}
where $<>_{f,s}$ indicates the effective Hamiltonian used for the calculation of the 
averages. We shall denote the nearest neighbor and next-nearest neighbor values of $Q_{ij}$
as Q and $Q'$ respectively.

We are thus left with two coupled Hamiltonians: a renormalized free fermions Hamiltonian 
for the $f_{i\sigma}$ and a lattice hamiltonian for the slave-spins that retains the full 
complexity of the original problem. We have thus to perform a further approximation, in 
this case the cluster mean-field on the spin Hamiltonian.

\begin{figure}
\includegraphics[scale=0.5]{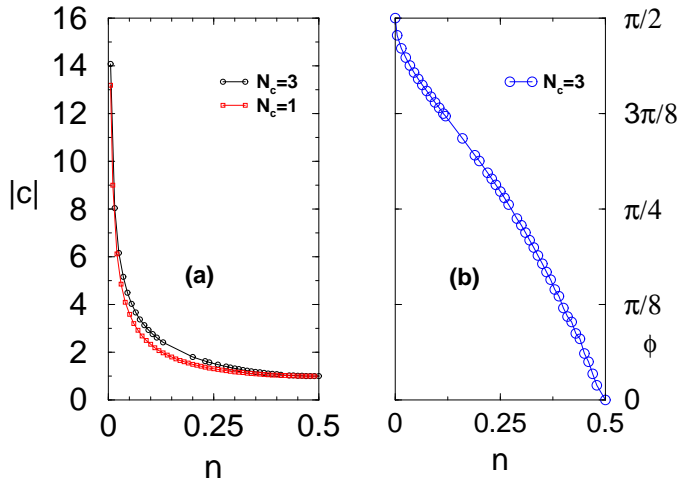}
\caption{Modulus and phase of the gauge $c$ that determines the choice of the 
proper hopping operators in the enlarged Hilbert space, in order for the CMFA to 
reproduce the non-interacting limit.}
\label{fig:c_phi}
\end{figure}

A cluster with a finite number of sites only is considered, within which 
interactions are treated exactly, and is embedded in the effective ("Weiss") field 
of its surroundings. A tiling of the original lattice is made, out of copies of the 
chosen cluster unit (Cluster shapes are chosen to respect lattice symmetry.), assuming 
translational invariance in the superlattice defined by this tiling, and this approximate 
Hamiltonian is used to calculate the mean-field average values.

In practice this means that in this approximation an effective Hamiltonian of a finite cluster 
is enough to represent the physics of the full lattice and that the "Weiss fields" are calculated 
using this same Hamiltonian (i.e. self-consistently) that represents also the surroundings of the 
cluster unit and not only the cluster unit itself.

Mathematically, we consider the following Hamiltonian for the slave-spin cluster $\it C$ :
\begin{equation}
H^{\it C}_{s}=\sum_{<ij>\epsilon {\it C}}{H}_s [i,j] +\sum_{i\epsilon {\it C},\sigma}h_{i\sigma}O^\dagger_{i\sigma}+ H.c.+\sum_{i\epsilon {\it C}}h^{z}_{i}S^{z}_{i},
\end{equation}
where  $h_{i\sigma}$ and $h^{z}_{i}$ are the effective fields of 
the surroundings, that are determined by the following self-consistency conditions:
\begin{equation}
h_{i\sigma}=\sum^{\prime}_{j n.n. i}J_{ij}\langle O_{i\sigma}\rangle
\label{eq:h}
\end{equation}
\begin{equation}
h^{z}_{i}=\sum^{\prime}_{j n.n. i}V \langle S^{z}_{j\sigma}\rangle
\label{eq:hz}
\end{equation}
where the prime over the sum means that sites j inside the cluster are excluded. We solve the spin 
Hamiltonian on cluster size $N_c=3$ for the triangular lattice and $N_c=4$ for the square lattice.     

It is useful to underline the role of two key quantities, in characterizing the physics of the system. 
It can be shown that $Z=<O_{i\sigma}>^2$ is the quasiparticle weight, while the effective mass enhancement 
is set by the effective hopping renormalization $Q_{ij}$. The two quantities coincide if the mean-field 
approximation on the slave spin hamiltonian is taken at the single-site level. Thus, they both vanish in the Mott
insulating phase. This amounts to neglecting all number fluctuations within the Mott phase. This is too
crude of an approximation especially when close to the Mott transition. On the contrary in the CMFA that we 
consider here these two quantities are distinct and one can have e.g. a Mott transition where the mass stays 
finite as we will see in the following.

We use Z as an order parameter: $ Z\ne 0$ indicates a metallic state, while Mott/CDW 
insulating behaviour corresponds to $Z=0$. 

\subsection{Choice of the gauge $c_{i\sigma}$}

We now discuss how to fix the gauge represented by the complex number $c_{i\sigma}$. 

The physical condition that we choose to impose is that our CMFA reproduces correctly 
the non-interacting limit, i.e. when $U=V=0$, 
\begin{equation}
Q_{ij}=Z=1,
\label{eq:nonint} 
\end{equation}
for any given filling $n^f$, so that $c_{i\sigma}=c(n^f_{i\sigma})$.

\begin{figure}
\includegraphics[scale=0.5]{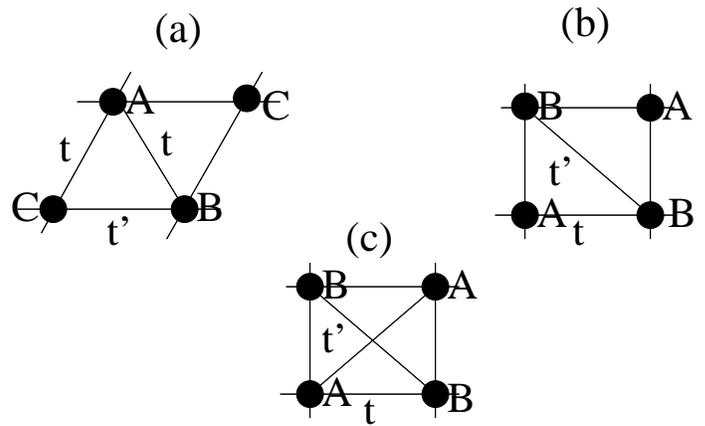}
\caption{Illustration of the lattices with hopping amplitude
$t$ and $t'$. (a) $\sqrt{3}\times\sqrt{3}$ sublattice decomposition of the
anistropic triangular lattice (ATL) (b) and (c) two sublattice decomposition
of the anistropic frustrated square lattice (AFSL) and the isotropic frustrated 
square lattice (IFSL). A, B, and C indicate the sublattice decomposition.}
\end{figure}
In the single-site approximation c can be chosen purely real and it can be determined 
analytically (as detailed in Appendix A). It takes the form:
\begin{equation}
c=\frac{1}{\sqrt{n(1-n)}}-1
\end{equation}

More generally $c$ has to be determined numerically by solving the mean-field equations 
at $U=V=0$ and imposing the conditions (\ref{eq:nonint}) and is a complex number, i.e. $c=|c|e^{i\phi}$ 
In Fig. \ref{fig:c_phi} we show $|c|$ and $\phi$ as a function of $n^f$ for a triangular cluster and for 
the single-site result, both on a triangular lattice.
 
We note that in both cases at half-filling $\alpha=1$ and $\phi=0$, and $O_{i\sigma}$ coincides 
with the form chosen in Ref\cite{Luca}, as anticipated.

\section{Hubbard Model}
The Mott transition, i.e, the metal-insulator tranisition driven
by the strength of electron-electron interaction in a homogenous
phase, has been studied in a great detail using various approaches
such as slave bosons, DMFT and its extensions.
In this section, we revisit the Mott transition on the lattices
shown in Fig.1. The control parameters are interaction strength
U/t and frustration strength $t'/t$, the ratio of next nearest neighbor to 
nearest neighbor hopping amplitude. As a function of these parameter, 
the Hubbard Model at half filling has, within CMFA, four possible
phases : a paramagnetic metallic phase, a paramagnetic insulating
phase, and insulating antiferromagnetic phase, and (in the 
presence of frustration ) an itinerant antiferromagnetic phase. However,
we shall be concerned here with the transition between the paramagnetic
metal to parmagnetic insulator. We study the paramagnetic solution 
by enforcing the spin symmetry  hence  avoiding the opening of
a full spectral gap due entirely to magnetic ordering.

The single-site cluster $N_c$=1 mean-field theory of slave spin representation 
gives the same results of the Gutzwiller approximation. In this regards CMFA provides 
a way to go beyond the Gutzwiller approximation. 

\begin{figure}
\includegraphics[scale=0.4]{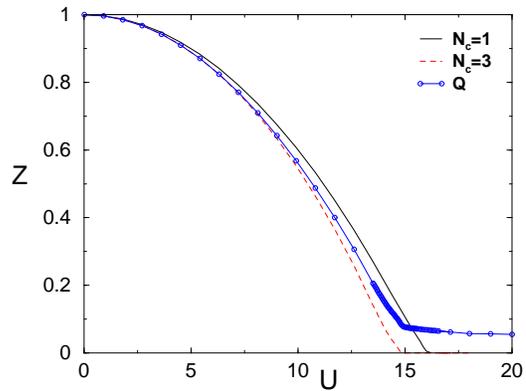}
\caption{ The order parameter $Z$ and the effective hopping $Q$ as
function of $U$. $U$ is measured in unit of $t$.}
\label{fig:op_vs_U_1}
\end{figure}

 First, we discuss the Mott transition on the isotropic triangular lattice. 
To get the uniform phase solution, we enforce the Lagrange multiplier $\lambda_{i}$   
and complex number $c$ to be the same for every site within the cluster. 
In Fig.\ref{fig:op_vs_U_1}, we plot Z and Q as a function of $U$ 
at x=0 for cluster sizes $N_{c}=1,3$. The critical value $U/t$, at which the Mott 
insulating phase occurs is 16.2, in the single-site ($N_c=1$) approximation, 
while it is around 15.1 in the three sites ($N_c=3$) CMFA. The short range 
correlations, which are built in the CMFA, supress the critical value $U$ by $6\%$. 
The critical value of U obtained from other methods such as DMFT-exact 
diagonalisation (8 site)\cite{Powell}, exact diagonalisation calculation 
for 12 site clusters \cite{Capone}, and cluster-DMFT(CDMFT) \cite{Bumsoo} 
are 15, 12 and 10.5 respectively. In CDMFT the tranisition is first order.
For $N_c=1$, the slave spin approach is identical 
to the Gutzwiller approximation \cite{Luca}, Q and Z are identical and they 
should vanish at the same critical value of U. We show, for $N_c=3$, Q 
as a function of $U$. It can be seen that it continues to be non-zero in 
the Mott insulating phase, and behaves as a $t/U$, as expected from the fact 
that the average kinetic energy is non-zero in a Mott insulator.

    It should be noted there is a substancial difference 
between $U_c$ obtained from 3-site slave-spin CMFA and 3- 
site CDMFT. It is because of CDMFT captures the fermionic 
quantum dynamics more accurately.

\begin{figure}
\includegraphics[scale=0.4]{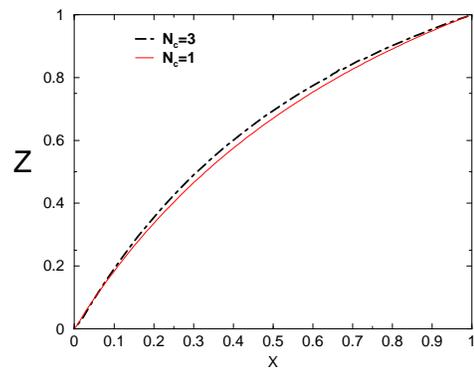}
\caption{The order paramter $Z$ as a function of $x$ in the large $U$ limit for $N_c$=1,3.}
\label{fig:op_vs_x}
\end{figure}

   We now examine $Z$ as a function of dopping in 
the limit $U\rightarrow \infty$ since this quantity
can be obtained in closed form in the Gutzwiller 
approximation. We show in Appendix B that for cluster size  $N_{c}$=1,
one recovers precisely the Gutzwiller approximation result 
$Z=2x/(1+x)$. 
In CMFA, we can ask how Z is modified in the  presence of short-range 
correlation effects. It can be seen in Fig.\ref{fig:op_vs_x} that the short range 
correlation effect on Z is appreciable for moderate to large doping 
$x$ and enhances Z in comparison to cluster size $N_{c}=1$. 

\begin{figure}
\includegraphics[scale=0.4]{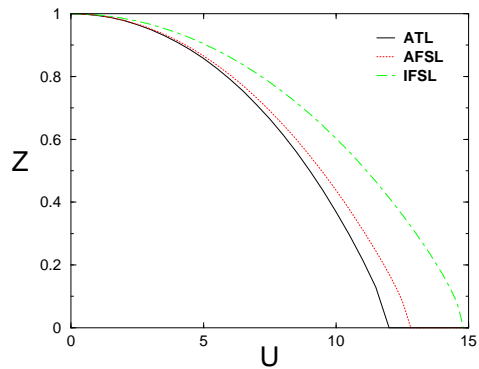}
\caption{ The order parameter $Z$ as
function of $U$ at $t'=0.4$ for the lattices ATL, AFSL, and IFSL. $U$ is measured in units of t.}
\label{fig:op_vs_U_ATL_IFSL_AFSL_tp_0.4}
\end{figure}

\begin{figure}
\includegraphics[scale=0.4]{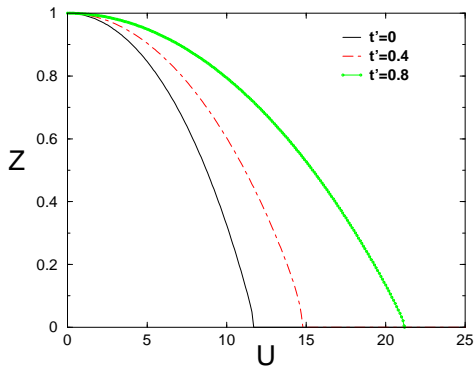}
\caption{ The order parameter $Z$ as a function of $U$ for various values of $t'$ for IFSL. $U$ is measured in units of $t$.}
\label{fig:op_vs_U_IFSL_various_tp}
\end{figure}

\begin{figure}
\includegraphics[scale=0.4]{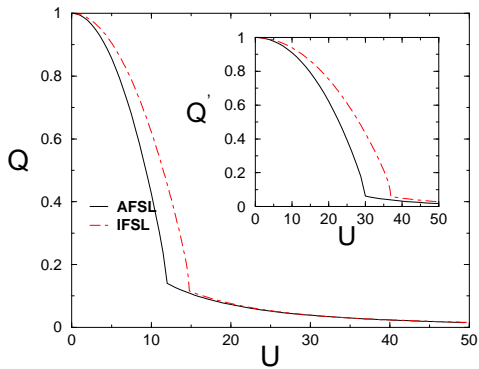}
\caption{ The nearest neighbor effective hopping $Q$ as
function of $U$ at $t'=0.4$ for AFSL and IFSL lattices. $U$ is measured in units of t. 
Inset shows the next nearest neighbor $Q'$ as a function $U$ in unit of $t'$ at $ t=2.5$. }
\label{fig:Q_Qp_vs_U}
\end{figure}
  
We now move on to the the dependence of the Mott transition on 
lattice and frustration.
In the absence of magnetic frustration on a bipartite lattice, 
one expects to find an antiferromagnetic ground state at low temperature. 
Ideally, the Mott transition can occur in system where antiferromagnetic
correlations are frustrated. In the $t-t'$ Hubbard model on the square lattice, 
a next-nearest neighbor hopping $t'$ frustrates antiferromagnetic correlations.  
By studying the lattices shown in Fig.2, we thus investigate the effects of frustration 
on the Mott transition in the half-filled $t-t'$ Hubbard model. For $t'=0$ the lattices 
shown in Fig.1 correspond to the unfrustrated systems and the effect of the frustation 
can be systematically studied as $t'$ is increased to its maximal value $t'=t$. 
Fig.\ref{fig:op_vs_U_ATL_IFSL_AFSL_tp_0.4} displays the order parameter $Z$ or 
single particle weight as a function of U at $t'=0$ for various lattices. 
At this value of $t'$, the critical value $U$ for the Mott transition is the 
lowest on the anistropic triangular lattice (ATL), while it is the highest on  
the isotropic frustrated square lattice (IFSL). As the  frustration $t'$ increases, 
the critical value of the Mott transition increases as shown in Fig.\ref{fig:op_vs_U_IFSL_various_tp}
for the isotropic frustrated square lattice. This increase with 
$t'/t$ is also seen in the Variational Cluster Approximation\cite{Andryi}. 
In Fig.\ref{fig:Q_Qp_vs_U}, we show the nearest and the next-nearest neighbor 
effective hopping $Q$ and $Q'$ of auxilary fermions. It can be seen that deep in 
the insulating phase they behave as $t/U$ and $t'/U$ respectively. Non-zero values 
of $Q$ and $Q'$ in the insulating phase signal that auxilary electrons (not the physical 
electrons) have a Fermi surface (with Luttinger Volume). It also implies, in contrast 
with infinite dimension (where single-site mean-field theory is exact), that in finite 
dimension the effective mass does not diverge in the insulating phase, despite the 
fact that $Z\rightarrow 0$. 

\begin{figure}
\includegraphics[scale=0.4]{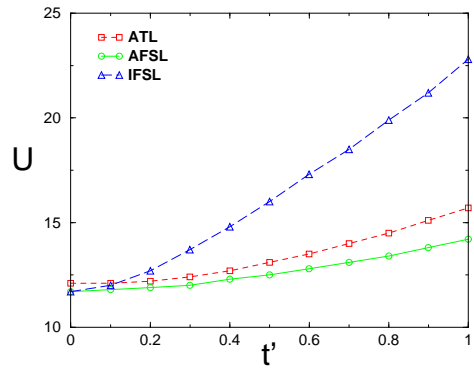}
\caption{ The order parameter $Z$ and the effective hopping $Q$ as
function of $U$. $U$ and $t'$ are  measured in units of $t$.}
\label{fig:op_vs_U}
\end{figure}

     Finally, we show in Fig.8 the phase diagram in $U-t$' plane for the above mentioned three lattices. 
One notices that the maximally frustated lattice, the triangular lattice, has the lowest
critical value of the Mott transition, while isotropic frustrated square lattice has the highest 
critical $U$.


\begin{figure}
\includegraphics[scale=0.4]{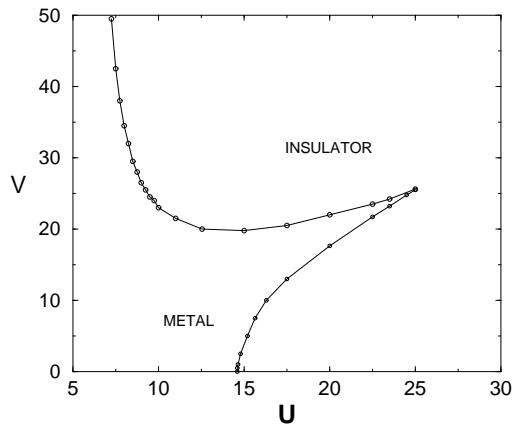}
\caption{This phase diagram displays critical values of $U$ and $V$ 
where $Z$ vanishes at half-filling. $U$ and $V$ are measured in unit of $t$. }
\label{fig:fig_n_0.5}
\end{figure}



\begin{figure}
\includegraphics[scale=0.4]{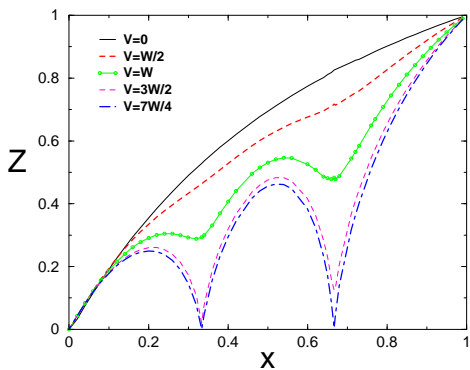}
\caption{$Z$ as a function of dopping $x$ at $U=100t$ for various values of 
$V$. W is the full bandwidth of the isotropic triangular lattice (W=9t). }
\label{fig:Z_vs_x_at_U_50}
\end{figure}

\section{The extended Hubbard Model}
\subsection{Uniform Phase}
In this section, we consider the extended Hubbard model on the 
isotropic triangular lattice for different ranges of parameters 
$U$, $V$, and doping ($x$) in the uniform phase, by enforcing that the Lagrange 
multiplier be the same at every site on the cluster and thus avoiding charge ordering.
Let us first examinethe combined effect of $U$ and $V$ on $Z$ at $x=0$. For given U
or V, we compute the crititcal value of U or V at which the Mott transition
occurs.  This study leads to the uniform ground state phase diagram
in the $U-V$ plane that is shown in Fig. \ref{fig:fig_n_0.5}. 
For $U<7.5$ the system is in the metallic state for any values of $V$.
For $7.5<U<15$, the system enters into the Mott insulating phase upon increasing V. 
We note however that there is a 'reentrant' structure of the metallic phase at still larger
V. This 'reentrant' structure emerges when $U$ and $V$ are 
comparable. It is because of $V$ compensates the effect of $U$.
And moving a nearest-neighbor to have a doubly occupied site, as in a
metallic phase, may become energetically favourable since the repulsion on the 
nearest-neighbor is comparable to that on-site. 

 From study of Sodium Cobalt Oxide in Ref.\cite{Chou}, it
appears that there is a large supression of the valence-band 
width$\--$by an order of magnitude compared with the local 
density approximation (LDA) band structure calculation\cite{LDA}. 
Ref.\cite{Motrunich1}suggested that such large renormalization of 
the hopping may be caused by $V$, and thus $Q$ was studied 
as a function of $x$ for different values of $V$ by means of the 
Jastrow-Gutzwiller (JG) wave function. The $Q$ in JG wave 
function study is equivalent to Z in our case. Slave-spin CMFA 
should be more accurate than the Jastrow-Gutzwiller approximation since, 
it captures the short-range correlation effect of $V$ in a better way 
because this term is treated exactly on the cluster.


In Fig.\ref{fig:Z_vs_x_at_U_50}, we show $Z$ as a 
function of $x$ for different values of $V$ at ($U=100$).
It should be noted that the value of Z vanishes at the commensurate dopings
$x=1/3$ and $x=2/3$ when $V$ takes its largest value, $V=7W/4$ (where $W=9t$ is the full
bandwidth of the isotropic triangular lattice). At doping $2/3$ the dominant configurations
at large $V$ on any triangle are ($\downarrow$, $\downarrow$, $\uparrow$) ($\downarrow$, 
$\uparrow$, $\downarrow$) ($\uparrow$, $\downarrow$, $\downarrow$). Now Z involves filipping a spin. So we have to make transition to states
like (($\downarrow$, $\downarrow$, $\downarrow$) or ($\uparrow$, $\uparrow$, $\downarrow$) etc. These have a higher
energy in the presence of V. Similar arguements holds for at doping $1/3$. Z vanishes in our case around $V=7W/4$, which is quite a large value in comparsion to the JG study, where it occurs at $V\simeq W$. This implies that JG study
overestimtes the effect of the short-range correlation of V.

     The effect of $V$ on the effective hopping $tQ$ is not as strong as we observe 
on $Z$ (not shown) since on a single triangle, there is no cost to move the particle 
via a kinetic move, eg: 
($\downarrow$, $\downarrow$, $\uparrow$) $\rightarrow$ ($\downarrow$, $\uparrow$, $\downarrow$).
It is true on a single triangle not connected on anything else, but the mean-fields connected
to the triangle will have small effect, which manifests  itself by a small supression in the 
effective hopping. Going beyond a single triangle however, it is clear that the effective hopping (bandwidth)
will be suppressed, a possiblity which we do not consider here.

\subsection{The CDW instability}
  In this section, we study the instability of a non-ordered 
phase toward a CDW in the presence of V. We determine for a few dopings 
the ground state phase diagram in  the $U-V$ plane of the system that 
has the $\sqrt{3}\times\sqrt{3}$ 
ordering pattern. We solve the above mentioned equations allowing for site-dependent Lagrange multipliers.
The procedure is as follows: We allow the complex number gauge c 
and the Lagrange multiplier $\lambda$ different for each sublattice and Q's and J's 
are different for every bond on the cluster.  We fit the filling dependence
of the magnitude of the complex number c and its phase to obtain c for
different sublattices, based on the c(n) relation obtained on the noninteracting system..

\begin{figure}
\includegraphics[scale=0.4]{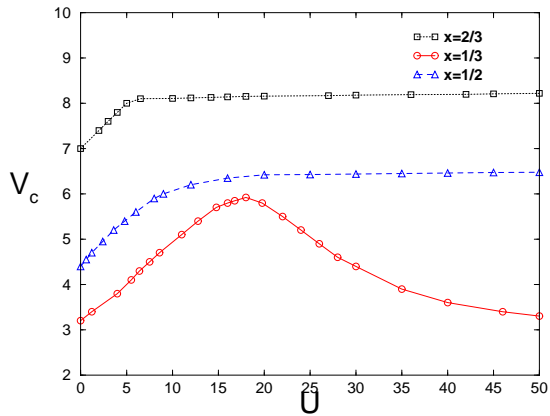}
\caption{Phase diagram in $U-V$ plane for $x=1/3,1/2,2/3$. The CDW phase 
is above the lines for the corresponding fillings.}
\label{fig:CDW_phase_diagram_U_V}
\end{figure}

Fig.(\ref{fig:CDW_phase_diagram_U_V}) shows the resulting CDW phase diagram in 
the  $U-V$ plane for three special values of $x$. 
The transition from metallic to  
CDW phase is first order. We also note that the effective mass $1/Q$ diverges at the transition.
We note that the lowest value of $V_c$ is at $x=1/3$. We also do not 
find that dopings $x=1/3$ and $2/3$  are playing any special role, as 
was suggested in the JG study. It should also be  noted that we 
find the CDW state at $x=0.5$(not shown)  in contrast with the prediction of the 
uniform phase in our study (where Z never vanishes) and in the JG study\cite{Motrunich1}.
The slave-boson mean field study of Ref\cite{Motrunich2} also
predicts a phase diagram similar to ours. However, our method 
captures the short-range effect of correlations in a better 
way than Ref\cite{Motrunich2}. We suspect that our phase 
diagram does not match with the diagram proposed by the JG 
study because it underestimates the effect of $V$.
 
\section{Conclusion}
 We have presented an extension of the slave-spin formalism away from 
half-filling by introducing a gauge variable. And we have shown how to 
solve the resulting model in the cluster mean-field approximation (CMFA). 
While in the single-site mean-field approximation the gauge variable can be chosen as pure 
real number, it is a complex number in the cluster approximation. 
This number changes from a pure real to a pure imaginary number as we move from 
the haff-filled to the empty lattice. The advantage of this method lies
in the fact that the short-range correlations can be  properly
taken into account. In the single-site approximation for the Hubbard model, 
we found analytically that in the infinite U limit, the single-particle weight Z 
reproduces the Gutzwiller result. In the CMFA, short-range correlations modify this result. 
The modifications are more important for intermediate dopings but they are never very large.

We have applied this approach to the Hubbard and to the extended Hubbard Model. 
In the case of the half-filled Hubbard Model, we have revisited the Mott transition on 
three class of lattices: anistropic triangular lattice (ATL), 
High-T$_{c}$ lattice(ISFL), and organic superconductor lattice (AFSL). We have done a 
detailed study of the critical value $U$ where the Mott transition occurs 
as a function of the frustration strength $t'$, and have shown that the effect 
of $t'$ in the presence of the short range correlations  is to increase the critical value 
for the Mott transition $U_c(t')$.

  We have also studied the extended Hubbard model in two 
dimensions in the uniform phase and shown that there is a reentrant 
structure between the insulating and metallic phases when $U$ and $V$ 
are comparable. We have also shown that dopings 1/3 and 2/3 play a special
role in the uniform phase. The quasiparticle weight can vanish at these
dopings. 

   For the extended Hubbard model, we have found two ground state 
phases on the triangular lattice: the metallic and the  $\sqrt{3}\times\sqrt{3}$ 
CDW state in a broad doping regime. At the present level 
of approximation, we found that, contrary to the uniform phase, 
dopings $1/3$ and $2/3$ in the CDW state do not play a special role.

     Finally, we point out that this method can be used to study magnetic 
phases. That has been left for future work. It can also be applied to study 
the physics of the multiband Hubbard model away from half-filling and can be 
generalized to tackle the t-J Model, and other strongly correlated models.

\section{Acknowlegdments}
The authors thank A. Georges, O.Parcollet, and S. Florens for discussion 
at the initial stage of the project and Ecole Polytechnique where this work was begun. 
SRH is greateful to A.-M.S Tremblay 
for important suggestions and many insightful discussions. SRH also thanks
A. Paramekanti for useful conversation. Computations were performed on the Dell 
cluster of the RQCHP. The present work was supported by NSERC (Canada), and the 
Center for Materials Theory, Rutgers University (LdM).

\appendix
\section{Choice of the gauge c in the single-site mean-field}
\label{app:c_single-site}

In the single-site approximation, we can determine the gauge $c$ analytically\cite{Luca_thesis}. 

The non-interacting single-site slave spin Hamiltonian $H_{s}$ reads\cite{Luca} 
\begin{equation}
H_{s}=hO^{\dag} + h^*O +\lambda (S^z+\frac{1}{2}),
\end{equation}
where $O$ is defined as in eq. (\ref{eq:O}). The single-site fermion part of the Hamiltonian is simply spinless non-interacting fermions.
The physical spin index $\sigma$ is supressed in $H_{s}$ since for $U=0$ upspin and downspin fermions are  decoupled, so that we can diagonalize the hamiltonian for one slave-spin in the $S^z=\pm 1/2$ basis.
The ground state eigenvalue $\epsilon_{GS}$ and the corresponding eigenstate are
\begin{equation}
\epsilon_{GS}=-\sqrt{\frac{\lambda^2}{4}+\vert a^2\vert}\equiv-R
\end{equation} 
\begin{equation}
\vert GS\rangle=\left(\begin{array}{c}\frac{\frac{\lambda}{2}+R}{N}\\\frac{-a*}{N}\end{array}\right)
\end{equation}
with $N=\sqrt{2R(\frac{\lambda}{2}+R)}$ and $a=h+ch^*$.

The expectation value of $S^{z}$ and $O$ in the ground state are 
\begin{equation}
\langle S_z\rangle= \frac{\lambda}{4R}
\end{equation}
and
\begin{equation}
\langle O \rangle=-\frac{ca^*+a}{2R}
\end{equation}
The Lagrange multiplier depends on the density $n$ and is adjusted in order to satisfy the constraint equation: 
\begin{equation}
n-\frac{1}{2} = \langle S_z\rangle=\frac{\lambda}{4R}
\label{eq:n}
\end{equation}

We want to tune $c$ in order to match the condition that in the limit $U=0$ the renormalization factor Z must be unity:
\begin{equation}
Z=<O>^2=\frac{\vert c a^{*}+a\vert^2}{4R^2}=1
\label{eq:Z_1}
\end{equation}

We can easily eliminate $\lambda$ from these two conditions, by squaring eq. (\ref{eq:n}).
We are left with the following expression for $c$:
\begin{equation}
\frac{\vert a \vert^2}{\vert c a^*+a\vert^2}=n-n^2
\end{equation}

If we choose c to be real then h and a are also real.
Then, the expression for $c$ in the closed form is
\begin{equation}
c=\frac{1}{\sqrt{n(1-n)}}-1.
\end{equation}
Note that this result is independent of $h$.

This cannot be done in the cluster case, since also the condition Q=1 has to be imposed 
and c has to be chosen complex in order to satisfy this further equation.

\section{Slave-spin formulation of the infinite-$U$ limit of the Hubbard model}

We derive here the analytic expression for $Z$ as a function of doping in the infinite $U$ limit and in the single-site approximation.

In this limit, no double occupancy is allowed so that the interaction term is replaced by a projector that enforces this constraint.
In order to do this we replace $O_{i\sigma}$, as defined as in eq. (\ref{eq:O}), by
\begin{equation}
{\tilde O}_{i\sigma}=(\frac{1}{2}-S^{z}_{i\bar{\sigma}})O_{i\sigma}
\end{equation}
where $\bar{\sigma}=-\sigma$. We thus obtain, for  the single-site mean-field spin Hamiltonian $H_{s}$:
\begin{equation}
H_{s}=\sum_{\sigma}h_{\sigma}{\tilde O}_{\sigma}^{\dag} + H.c.+ \lambda \sum_{\sigma}(S^z_{\sigma}+\frac{1}{2})
\end{equation}
with 
\begin{equation}
h_{\sigma} = -\sum_{j}J_{ij}\langle \tilde O_{j\sigma}\rangle
\end{equation}
where $j$ indicates the neighbor of site $i$. Diagonalizing $H_{s}$, we obtain the ground state eigenvalue and eigenvector, i.e.
\begin{equation}
\epsilon_{GS}=\frac{\lambda}{2}-\frac{1}{2}\sqrt{\lambda^2+8\vert a \vert^2}
\end{equation}

\begin{equation}
\vert GS \rangle = \left(\begin{array}{c} 0\\a/N\\a/N\\\epsilon_{GS} \end{array}\right)
\end{equation}
with 
\begin{equation}
N=\sqrt{\epsilon_{G.S}^{2}+2\vert a \vert^2}
\end{equation}

Hence we can determine $\langle S^z_{\sigma}\rangle$ and $\langle O^{\dag}\rangle$.
\begin{align}
\langle S^z_{\sigma}\rangle&=-\frac{1}{2}\frac{\epsilon_{G.S}^2}{N^2} \nonumber \\
\langle O^{\dag}\rangle&= \frac{a^* \epsilon_{G.S}}{\epsilon_{GS}^2 + 2 \vert a\vert^2} (1+c).
\end{align}
The Lagrange multiplier is fixed by the constraint equation that depends on the chosen filling:
\begin{equation}
n=\langle S^z \rangle + \frac{1}{2}=\frac{\vert a \vert^2}{\epsilon_{GS}^2+2\vert a \vert^2}
\end{equation}

We can calculate the renormalization factor  $Z=\vert \langle O^{\dag}\rangle \vert^2$ using 
\begin{eqnarray}
Z&=&\vert \langle O\rangle \vert^2=\vert 1 + c\vert^2 \frac{\epsilon_{GS}^2}{\vert a \vert^2}n^2\nonumber \\
&=& \vert 1 + c\vert^2 n(1-2n)
\end{eqnarray}

Using the one-band prescription (see appendix \ref{app:c_single-site}) $c=\frac{1}{\sqrt{n(1-n)}}-1$, 
we then obtain
\begin{equation}
Z=\frac{1-2n}{1-n}=\frac{2x}{1+x},
\end{equation}
 (where $x$ is the total doping, $2n=1-x$). That is precisely the result of the Gutzwiller approximation.

\end{document}